# Calcium imaging and analysis of the mouse hippocampus or neocortex using miniature microendoscopes

**Authors:** Jessica Winne[1,2], George Nascimento[2], Ingrid Nogueira[1], Richardson N. Leão[2] & Katarina E. Leão[1,*]

**Author Affiliation**

1. Hearing and neuronal activity lab, Brain Institute, Federal University of Rio Grande do Norte, postal box 1524, Campus Universitário Lagoa Nova, CEP 59078-970, Natal/RN, Brazil

2. Neurodynamics lab, Brain Institute, Federal University of Rio Grande do Norte, postal box 1524, Campus Universitário Lagoa Nova, CEP 59078-970, Natal/RN, Brazil

**Author contribution:**

**Jessica Winne:** Methodology, Data curation, Formal analysis, Visualization, Writing- Reviewing and Editing. **George Nascimento:** Methodology, Software, Formal analysis, Visualization, Writing- Reviewing and Editing. **Ingrid Nogueira:** Validation. **Richardson N. Leao:** Methodology, Formal analysis, Visualization. **Katarina E Leao:** Writing- Original draft preparation, Visualization, Conceptualization, Writing- Reviewing and Editing.

**\* Corresponding author**
Katarina E. Leao: katarina.leao@neuro.ufrn.br

Brain Institute, Federal University of Rio Grande do Norte, postal box 1524, Campus Universitário Lagoa Nova, CEP 59078-970, Natal/RN, Brazil

**Running head**: Miniscope imaging hippocampus and neocortex

**Funding sources:** This work is supported by the American Tinnitus Association and the Brazilian National Council for Scientific and Technological Development

**Conflict of interest:** The authors declare no conflict of interest

**Abstract**
Calcium imaging using miniscopes are becoming increasingly popular in the neuroscience community with a multitude of microendoscope versions, lens systems and packages for analysis available as open source. Here we describe in detail how to implant lenses and baseplates for the University of California at Los Angeles (UCLA) miniscope V3 and V4 using a Gradient Index (GRIN) lens for the hippocampus or a combination of prism and GRIN lens for recording in the neocortex. Our protocols contain adaptations from several published protocols however the GRIN-relay-Prism lens system protocol is developed in-house and not previously described. This chapter also suggests how lenses may be re-used and how to test the quality of a reused lens. Lastly, we comment on the difference between analysis software and the computational needs for different packages.

# 1. Introduction

Miniature microendoscopes, or miniscopes, used for imaging calcium transients from genetically encoded calcium indicators (GECIs) are becoming increasingly popular for studying neuronal activity during behavior in small animals like mice or birds. The miniscopes are headfixed devices, lightweight, requiring only implantation of a lens and cyanoacrylate glue securing a baseplate whereto the miniscope is secured during awake, freely moving behavioral tasks. The miniscope allows for tracking activity of hundreds of neurons in one go. Furthermore, analysis refinements make it possible to identify and observe specific neurons over extended periods of time. Another advantage of the miniscope is sampling of single-cell activity from a relatively large region of neuronal tissue at the same time and during different behavioral tasks. With the advances of fast GECIs[1], micro optics[2], lightweight wireless devices[3], and powerful analysis programs for single cell tracking[4,5] it is possible to study the dynamics of intracellular calcium activity close to or at single spike resolution[1,6]. Here we describe a brief summary of current types of miniscopes and next provide a protocol for implanting different types of lenses in the hippocampus and neocortex, and the securing of the baseplates for optimal imaging using miniature microendoscopes.

### 1.1 Miniature microendoscopes
The first miniature microscope with an epi fluorescent (a parallel beam of light passed directly upwards through the sample, maximizing the amount of illumination) light source was published in 2011[2] by the Mark Schnitzer group. This equipment, weighing only 1.9g was assembled from mass-produced pieces, including a semiconductor light source and a complementary metal oxide semiconductor (CMOS) sensor (which converts photons into electrons for digital processing) and could record calcium activity of hundreds of purkinje cells during different motor behaviors[2]. Now, other research groups have developed different models of mini microscopes with small variations between them in regards to weight (1.6 - 4.5g), material, one or two scopes, opto-remote (ability to optogenetically stimulate outside of the imaging field) and automatic focus[3]. Examples of some current models are the NINscope[3], University of California at Los Angeles (UCLA) Miniscope V3[7], UCLA wireless Miniscope[8], UCLA V4 Miniscope[9], FinchScope[10], compact head-mounted endoscope (CHEndoscope)[11], miniscope GRadient INdex (GRIN) lens system[12], wireless miniScope[13], bilateral microendoscopes[14], and lightweight miniscope for subcellular imaging[15]. There has also been development of portable two-photon microendoscope devices[16–18] to avoid downsides of single photon imaging such as photobleaching, phototoxicity, light penetration/light dispersion issues. Interestingly, the recently described miniature head mounted two-photon fiber-coupled microscope also allows for tilting the focal plane by 30 degrees to image functionally distinct neurons above and below regions of interest[16]. Still, here the focus is on 1-photon imaging of calcium activity, specifically using miniscope models developed at UCLA.

**1.2 UCLA Miniscope models (V3 and V4)**

The Miniscope open source project has been under development at UCLA since mid 2015[7]. Currently, this miniature microscope is the most widely used for neural recording in freely behaving animals, probably due to the UCLA Miniscope being open-source, customizable, affordable, and providing comprehensive technology with a large community of users actively exchanging information in open forums. About 500 laboratories worldwide are part of the Miniscope user community[19]. The V3 UCLA miniscope consists of a machined *delrin* (acetal homopolymer) housing, filters and optical lenses, an excitation light source and a CMOS image sensor. The weight of the system is less than 3 grams and the adjustment of the focal plane (~ 150μm) is achieved by manually adjusting the distance between the CMOS image sensor and other optical elements by adjusting the miniscope body length. A modification of the version 3.2 miniscope is the Wireless UCLA Miniscope[20]. In 2020, the V4 UCLA Miniscope was launched that presents advances in relation to the previous models such as 1mm diameter field of view and greater working distance (~1mm), electronic focal adjustment (+/-200 μm), all achromatic optics, lightweight (2.6 g) and a head orientation sensor and motion sensor. Additional improvements with the V4 miniscope is the need for less excitation power compared to previous systems (~1/5th) and updated acquisition software[9]. Practically it is also useful to know that prism lenses can be used without a relay lens in the V4 version, however we will describe protocols for lens implants appropriate for both the V3 and V4 UCLA miniscope in the Methods section.

**1.3 Microendoscopic lens types**

Miniature microendoscopes together with gradient refractive index (GRIN) microlenses allow optical access of neurons located in different brain regions, both superficial and deep[21]. However, when the area of interest is the neocortex, it is important to consider that to image supragranular and infragranual cortical layers different lenses are needed. When neurons of interest are located in the superficial cortical layers (layer 1-3), the GRIN lens (that has a cylindrical shape with flat end surfaces), can be implanted without removing the dura mater[22]. However, for the imaging of cells in the deeper cortical layers (layer 4 - 6) currently the use of the prism lens is the most appropriate method[23]. In general, when opting for the GRIN lens, it is necessary to aspirate any superficial layers above the area of interest [24]. Instead, the prism lens is implanted laterally to the cells of interest, with the light from the microscope exciting fluorescently labelled cells located along the image face of the prism probe[25], and thus the implanted prism lens minimally affects the horizontal organization of the cells to be imaged.

The choice of length and diameter of the lenses depends on the region of interest for the study, where 0.5mm, 1mm, 1.8mm, and 2mm diameters are commonly offered by companies such as GRINTEC and Go!Foton (with possibility to customize lenses). When opting for thin GRIN lenses (0.5 -1 mm) together with the V3 miniscope (UCLA) or Inscopix from the Ghosh group[26], it is necessary to use two GRIN lenses or one GRIN lens and the prism lens[21,25].

**1.4 Miniscope imaging of the auditory cortex of mice**

The use of prism lens for focusing on the auditory cortex, with its lateral position, vastly improves the accessibility of in vivo calcium imaging of this region[23], compared to two-photon/head-fixed approaches where the whole microscope stage is rotated for access in awake mice[27]. The tonotopicity of the auditory cortex has recently been remapped in mice using cranial windows and transgenic GFP-based calcium indicator expressing mice (such as GCaMP6s), together with wide-field and two-photon imaging[27]. Here, miniscopes and GRIN-prism lenses systems could be useful for extending such analysis by focusing on sound responses in certain cortical layers[28], or focusing on sparser groups of inhibitory interneurons using specific cortical markers[29] while allowing the animal to be freely behaving during auditory processing tasks. In a previous study from our group we showed that previous (1 week) noise exposure alters firing properties of subtypes of layer 5 pyramidal cells (L5 PC) of the mouse auditory cortex[28]. Using relay lens/prism probes and constrained non-negative matrix factorization (CNMF) analysis[30], we could track the same layer 5 cortical neurons before and after the noise-exposure session (1 week apart) and compare activity to the same stimuli. We could thereby identify distinct PCs decreasing or increasing firing frequency to the same stimuli before and after loud noise exposure and correlate data to *in vitro* results[28]. Together this shows how GECI and miniscope imaging can simplify functional analysis of areas and/or behaviors that otherwise might require large and costly 2-photon stationary setups with head fixed animals.

## 2. Materials
The following material is necessary for carrying out the method described:

- Basic surgical equipment (mouse stereotaxic frame, isoflurane vaporizer, surgical heat pad, stereo surgical microscope, light source, tools and a dental drill)
- Dissection knife and blades (size 11 blades for the incision of the prism, section 3.4.8)
- Two 200µl pipette tips (suction system/holder of lenses).
- Vacuum pump system
- Micro screws (stainless steel, 3mm long)
- 1ml syringe
- Cotton swabs
- Cyanoacrylate glue
- Silicon to cover the implanted lens (for example *Kwik-Sil*)
- GRIN-relay lens (if using miniscope v3)
- Prism probe lens
- Base plate for miniscope
- Miniscope device
- 5V power source
- USB type B cable
- Data acquisition hardware and computer with Miniscope controller software

## 3. Methods

### 3.1 Virus injection
Before implanting the GRIN lens, it will be needed to virally transduce a genetically encoded calcium indicator (for example GCaMP6f, 750µl, dilution 1:3 in saline). It is usually recommended to inject the virus ten days before the lens implant[28], however there are also possibilities of directly coating a lens with virus or using a GCaMP transgenic mouse (*see Note 1*). The following procedure was written for imaging the hippocampus. Additionally, a section 3.3 and 3.4 shows Modification of protocol for imaging of the neocortex. The expected time for completing one of these procedures is around 50 min.

### 3.2 Steps for Implanting the Gradient Index (GRIN) lens (hippocampus)

1. Weigh the animal (*see Note 2*) and anesthetize it with isoflurane (5% initially in a small chamber, then 1-2% during surgery, *see Note 3*) and transfer the mouse to a stereotaxic frame on a thermal heating pad. Cover eyes with lubricant eye gel to avoid drying of corneas, and keep the animal in isoflurane inhalation (1-2%) with a nose fitted mask.
2. Remove hair on the scalp by shaving it off with a razor blade. Next open the skin on the scalp with a long incision. The edge of the skin can be gently sutured and stretched open for better accessibility (**Fig. 1A**).
3. Apply hydrogen peroxide (15%) to remove the connective tissue and also de-attached the small muscles from the scalp in the neck region to be able to open the skin all the way to the back (*see Note 4*).
4. Drill three small holes, placed in a triangle around the area of interest, with a dental drill without perfusing the bone (the drill opening should match the threads for the chosen micro screws) and place three screws (pre-rinsed in a drop of alcohol and thereafter saline) in the hole and screw tight (~3 turns) (**Fig. 1B-C**). Micro screws will help secure the cyanoacrylate glue around the implanted lens, as well as, securing the cyanoacrylic glue holding the baseplate of the miniscope later on. Apply saline regularly over the skull bones to keep the area moist between steps, but drying before drilling is needed.
5. Align the head completely straight in the stereotaxic frame by comparing the dorsoventral (DV) coordinate differences between the lambda and the bregma, and in between the two hemispheres, by placing a standard syringe in the stereotaxic frame as a pointer, and use it to lower and touch the skull (aiming for <0.05mm difference).
6. Mark the area of where the lens will be implanted (use the microdrill to mark the center) and next make the craniotomy to open a 2 mm diameter (the size of the lens) window in the skull using a dental microdrill (**Fig. 1D**). Either drill a circle around the center area or use a micro drill with a 2mm diameter[21]. Use a bent needle tip to lift off the bone (**Fig. 1E**).
7. Examine the cranial hole with a stereo surgical microscope and gently remove any bone residues around the edges of the hole and remove the dura mater using a bent needle tip and cotton swabs. Drip saline solution and clean the area for the lens implantation as carefully as possible. Avoid causing bleeding. If any small bleeding appears try applying saline and dab with cotton swabs to avoid any coagulated blood on the lens (*see Note 5*).
8. With a bent and blunt needle connected through the tubing of a vacuum pump (**Fig. 1F**) gently aspire the cortex until reaching the corpus callosum which can be visualized as a slightly whiter color than the cortex using the stereo surgical microscope.

9. Position a 2x5mm lens inside a modified plastic pipette tip attached to the stereotaxic arm (**Fig. 1G**) using the vacuum suction system to hold the lens in place[31]. Then gently lower the lens into the cranial hole (**Fig. 1H**) and turn off the vacuum pump to set the lens in place, but do not remove the plastic pipette tip until the lens has been firmly secured by acrylic glue (*see Note 6*, step 10).
10. Remove any excess liquid with cotton swabs and apply cyanoacrylate glue between the lens and the screws, creating a fixed helmet on the bone. Use UV curing cyanoacrylate glue if possible for a fast procedure. In detail, apply a drop of glue with the bent needle, harden with UV-light, repeat and continue until the lens is secured to the skull and the lower part of the micro screws (*see Note 7*, **Fig. 1I**).
11. If the skin around the cyanoacrylate glue helmet is loose, suture with 4-0 nylon thread. Use two knots in the front and two knots in the back of the incision to close the skin around the helmet a little bit and apply silicon glue on top of the GRIN lens to protect it from any mechanical damage the animal might generate (**Fig. 1J**). Clean the skin around the incision with iodized soap or antiseptic solution, such as chlorhexidine.
12. Turn off the isoflurane flow (leave just air or oxygen flow on until the animal awakes). Remove the animal from the stereotaxic frame (post-surgery care in specific section below).

*Modification of protocol for imaging of the neocortex:*
### 3.3 Implanting the GRIN lens (L3 neocortex)
For imaging of the cortex layer 3 (L3) step 7 and 8 can be omitted. Imaging of L3 cortex does not require removal of dura mater nor any aspiration of the cortex. The working distance of the GRIN lens is sufficient to image L3 of the neocortex.

### 3.4 Implanting the Prism lens (L5 neocortex)
- Modifications to topics 6, 8, 9 and 11 of the hippocampus protocol:

**Topics changed below:**
6. Mark the area of where the lens will be implanted (*see Note 8*, **Fig. 2**) using a microdrill to mark the center) and next make the craniotomy to open a 1mm diameter (diameter of the prism lens) window in the skull using a dental microdrill.
8. Attach a straight-edged dissection knife to the stereotaxic apparatus holder arm and mount it so that the knife blade is perpendicular to the curvature of the skull. Lower the blade tip to 0.3mm above the area of interest and keep in place for 3 min (to ease the insertion of the prism lens triangular tip[25] in step 9).

9. Position the prism in place using a pipette tip adapted to attach the 1mm prism diameter (*see Note 9*). As the tip of the lens containing the prism has 1mm of length, the prism is lowered to 0.5mm below the DV coordinate of the area of interest (laterally to the prism, 0.3mm) as the image of the prism is in the center of the prism (see schematic drawing, **Fig. 2**).

11. Suture with 4-0 nylon thread at the incision site to close the skin over the lens and helmet (5-6 stitches). This procedure serves to protect the lens from any mechanical damage.

### 3.5 Post-surgery care

After surgery, the animal should be monitored until it returns to its normal behavior following local ethics protocol. During the 4 days post-surgery, treat the animal with analgesic and anti-inflammatory agents (*see Note 10*) according to local ethics prescriptions and rules. During this period, it is recommended to check daily the following parameters: weight, body temperature and if water and food are being ingested.

### 3.6 Attachment of baseplate

The baseplate will act as a docking station for the Miniscope device. Before placing the baseplate, it is useful to check for calcium signals as this will also help optimize the placement of the baseplate. Therefore, two weeks after the lens implant is a good time to perform this procedure, as it is enough time for the expression of the GCaMP, as well as for the reduction of blood clots in the imaged area (*See Note 11*).

### 3.7 Baseplate to GRIN lens

1. Attach the baseplate onto the bottom of the Miniscope (it has 3 magnets on each side) and a miniscope holder in the stereotaxic frame (*see Note 12*).
2. Anesthetize the mouse and place it securely in a stereotaxic frame.
3. Carefully remove the silicone lid from the top of the lens using curved forceps and clean the lens with isopropyl alcohol (used for cleaning objectives) with a cotton swab.
4. Using the stereotaxic apparatus, position the miniscope on top of the GRIN lens.
5. Explore the field of view of the miniscope by moving the stereotaxic arm holder in very small forward or backwards tilting angles (depending on what holder you are using). Small changes in angle can allow for visualization of optimal field of view. When identifying the region of interest, also adjust brightness of the LED (30% of max brightness[32]) and focus accordingly.

6. After finding your ideal view, fix the baseplate with cyanoacrylate glue to the anchor screws in the skull while the miniscope is still attached. Try to fix the baseplate in the best focal plane/position (thereby this will maximize the range that the microscope can be focused in either direction to pick up additional fluorescent signals).

### 3.8 Baseplate to GRIN-relay-Prism lens and V3 Miniscope
1. Attach a GRIN-relay lens (2 mm of diameter, 5 mm length) into the bottom of the Miniscope by holding the GRIN lens with fine forceps, applying a very small quantity of glue on the side of the GRIN-relay lens (*see Note 13*) and place it into the miniscope (laying upside down, **Fig. 3A** *see arrow*). Next you can focus the image on any object in the room (focusing the miniscope on something in the roof is practical since it is already laying upside down) to make sure the lens is positioned straight. If the lens has an angle the image will stay blurry and you have to redo the positioning. When the GRIN-relay lens is placed straight into the bottom of the miniscope, turn on the UV-light one time to harden the glue.
2. Place the miniscope in the upright position into a miniscope holder attached to the arm of the stereotaxic apparatus (**Fig. 3B**).
3. Anesthetize the mouse and place it securely in a stereotaxic frame.
4. Open the skin by cutting the thread of all stitches (5-6 stitches) with a fine pointed scissor. Carefully open the skin over the prism lens and clean the lens with Isopropyl alcohol using a cotton swab.
5. Using the stereotaxic device, position the miniscope with the GRIN-relay lens on top of the prism lens but leave a gap of air (~0.3mm air distance according to lens specification, but focus needs to be tested with the miniscope software, **Fig. 3C**).
6. Explore the field of view with the miniscope and adjust the LED (30% of max brightness[32]) brightness and focus accordingly (as in step 5 of the 'Baseplate to GRIN lens' section).
7. After finding your ideal view, attach the GRIN-relay lens above the prism lens with UV-hardened cyanoacrylate glue. Approximately 3mm of the GRIN-relay lens sticks out of the miniscope and needs to be firmly glued to the prism lens (*see Note 14*) without getting glue in the air space between the two lenses, only around them. Next the miniscope can gently be broken loose from the fused GRIN-relay lens/prism probe lens (**Fig. 3D-E**).
8. Next attach the base plate to the miniscope and lower it over the relay lens/prism probe (**Fig. 3F**). Explore the field of view with the Miniscope and after finding your ideal view, fix the baseplate with cyanoacrylate glue to the anchor screws in the skull while the miniscope is still attached (**Fig. 3G**). Try to fix the baseplate in the best focal plane/position (this will maximize the range the microscope can be focused in either direction to pick up additional fluorescent signals). Next the miniscope can be removed (**Fig. 3H**) with the baseplate and lenses firmly attached to the skull.

### 3.9 Baseplate to Prism lens and V4 Miniscope

For the V4 miniscope the GRIN-relay lens is not necessary, therefore following the topic 3.8, the steps 1 and 7 can be omitted. The base plate is then fixed similarly as described in step 8. The optimal focus determines the position for gluing the baseplate for the V4 miniscope.

### 3.10 Recording

Most miniscope recording sessions also require video recordings of the animal behavior, and usually this is done in a single computer by the miniscope software itself, or by a video camera capture software. For both cases it is advised to make tests and a careful inspection of consistency in synchronization as well as the total number of frames. Missed frames (due to a slow computer for example) are not acceptable as this cannot be corrected for after the recorded sessions (see Note 16).

To record data with the Miniscope (version 3 or 4) a data acquisition card/device (DAQ) is needed and a computer with Miniscope controller software installed or other optional software (*see Note 15*). At the beginning of each imaging session, the miniscope should be attached to the base plate (It will be necessary to habituate the animal for this procedure, otherwise the animal will need a quick isoflurane anesthesia for this step. In general, it is sufficient to habituate the mice by wearing a miniscope for 10min in the experimental room during at least two consecutive days, prior to the first recording session (**Fig. 4**). Focus of the miniscope is done manually for miniscope V3 (300 µm range) and digitally for V4 (+/- 200µm electronic focus). Optimum laser power (30% of max brightness[32]), imaging gain, and focal distance is selected for each animal and conserved across all sessions. To avoid the possibility of bleaching the fluorescence, beware of unnecessary long-lasting exposition of the laser during sessions.

The calcium signal can be acquired at a frame rate of 30 frames per seconds (or 20 frames/s if the fluorescence signal is low). The UCLA V3 Miniscope controller software allows for an automatic synchronization with a camera recording behavior, which records timestamps for each frame in a text file and provides I/O 3.3V TTL signals for synchronization with any external behavioral devices or other data acquisition systems[33].

### 3.11 Lens re-use

The implanted lenses are quite expensive but they may be possible to recycle for at least two or three times, depending on the process used for its recovery. For some experiments that require histological studies, forceps, pliers and scissors are used for brain dissection. Here, extra care should be taken to avoid touching these tools in the lens that is well cemented/glued in the skull. The lens borders are quite sensitive to mechanical stress and inevitably these operations will cause some cracks and breaks in the lens edges. However, gentile mechanical forces (using a plier to cut the bone/acrylate glue around the lens until the lens comes loose) can be employed to detach a lens from a fragmented skull. After cleaning with a pre-moistened lens cleaning wipe, lenses should be inspected in a microscope. Considering that the image registered from the miniscope camera is rectangular and the flat lens surface shape is circular, cracks of a few tiny areas in the borders may not interfere a lot in the recorded frames of image. Still, it is important to inspect the lens surface for scratches as a scratch in the center of the image will make the lens useless.

To inspect the lens quality one can, use a white paper marked with a fluorescent marker (i.e. yellow liquid marker pen), and paper glue to keep the lens standing upwards on the paper. Next, inspect the lens quality by visualizing the paper microscopic fibers as shown by the miniscope on the computer screen. If the mesh of fluorescent fibers becomes blurry at some location this can indicate a problem of the lens (dots or scratch on lens). Holding the miniscope in a stereotaxic frame may be useful for this step, as fine tuning the height of the miniscope will be necessary to focus. Any serious defect in the lens will show up in the image details as one fine tune the focus by moving above and below the focus point of paper fibers. To remove the paper glue, cleaning the lens with a lens cleaning wipe will be necessary. Be sure to remove all as any minimum amount of dye from the liquid marker pen produces fluorescence that could ruin experiments.

### 3.12 Analysis

For miniscope experiments there may be only a couple of percentages of change in fluorescence of an active neuron, also cells are seldom rigid in position due to movements of the animal. To detect specific neurons a process of motion correction is needed. This step is crucial and will also make it possible for following computations to separate the real signals from blood vessels, platelets, and reduce background from neurons outside the focal plane.

Processing of raw videos is still a challenge for many researchers using in vivo single-photon calcium imaging. Also, it demands high computing power and memory capacity, however for very large data sets it is possible to request a service on the cloud[34]. Calcium imaging processing basically consists of motion correction and cell extraction steps. If a certain cell or group of cells briefly shows any particular calcium dynamics of interest at a specific time during the recorded session, the algorithm has to take this particular cell(s) into account for the whole data set and at the end, the data has to be analyzed in a pixel by pixel basis. For the case of a recording session lasting 10 min, it is required about 6.2Mb for the raw data (for example 480x750pixles, 30frame/s), this data is converted to floating point format for processing and increases by a factor of four. To increase the computation speed, the data usually is down sampled by a factor of two in each spatial dimension and up to five in the temporal dimension but the quality of the final result can be degraded and needs to be evaluated. The data has to be fragmented in chunks of frames to be loaded in the RAM computer memory, processed in parallel and later the results combined. Luckily, for most of the packages the user just needs to feed the computer with the raw data, and set up a few processing parameters. The computer resource needs to have at least 16Gb of RAM memory and more than 8 cores, for a better parallel processing of these chunks of data[30]. It is highly advisable to keep the names of all miniscope generated raw data and folders as it was recorded, as some packages make use of this information for avoiding erroneous processing and to track the correct date and timing for the analysis. After extracting the calcium gradient from the raw data, another challenge in data processing has been to infer spike timing (spike rates in population imaging) from noisy traces. To solve this challenge, several different approaches have been used, such as template-matching, deconvolution, and approximate Bayesian inference[30,35–38].

With the popularity of using the miniscope, the analysis packages designed to detect, map, and extract the dynamics of calcium are growing[5,36,39]. The following details some of these methods, which were implemented as open source software and were designed initially to be fully compatible with the V3 Miniscope DAC software.

### 3.12.1 CalmAn

CalmAn is an open-source library for calcium imaging data analysis[40]. This package provides methods for solving problems ranging from preprocessing to registration across different sections of data that is required in analysis of large-scale calcium imaging data. CalmAn has versions in Python, and an older unsupported version implemented in Matlab[41]. The package is developed by the Flatiron Institute and is well tested, has a substantial amount of libraries, is well documented and has a large number of users. The theoretical background and the description of the parameters used in this package has the implemented methods based as NoRMCorre[42] for motion correction; CNMF constrained non-negative matrix factorization[30], CNMF-E[5], OASIS[43,44] for deconvolution and demixing of the imaging data and OnAcid[45], OnAcid-E[46] for online analysis.

The CalmAn package provides resources for running in batch, and also to be used from laptops to high end clustered computers. The efforts of many groups have led to designing several other very useful open source software packages for calcium imaging analysis. For example, for better compatibility in a limited processing computing one can use MINIAN[39] (*see Note 17*). For processing the raw videos in the cloud, PIMPN is an interesting package[47] (**Fig. 5**, *see Note 18*), and for registering of cells recorded in different sessions CellReg[48] is useful (*see Note 19*).

### 3.12.2 Miniscope graphics user interface (MESmerize)

All the packages cited so far provide their results in stored datasets, graphics and images. MESmerize[49] is a platform that was originally set to interact with the CaImAn package and based on visual interfaces it presents in a very handy fashion a preview of the raw calcium images. Mesmerize interactively gives the user controls for setting the parameters for motion correction, automated run batch programs in CaImAn and then check for the results in a process for optimization of those parameters. The same approach was designed for the next step, related to the extraction of the calcium dynamics. Mesmerize interactively provides the annotation and analysis for the entire process, from raw data to semi-final publication figures, and aids in the creation of FAIR-functionally linked datasets. It is applicable for a broad range of experiments and is intended to be used by users with and without a programming background. This package also provides an interesting way to correlate the results with the experimental stimulus paradigm, classification of cells and overlap the cell maps to brain atlas images.

## 4. Notes

**Note 1.** The UCLA miniscope forum[50] was recently discussing how to coat lenses directly with virus solution using fibroin, which is a silk protein[51] to reduce the number of surgical procedures in each animal. Furthermore, now many transgenic mice lines also express versions of genetically encoded calcium indicators such as the GFP-based GCaMP family and that can express GCaMPs in a Cre recombinase dependent fashion.

**Note 2.** Make sure the animal remains in good health throughout procedures and does not lose more than 10-15% body weight (according to local ethics recommendations).

**Note 3.** We find that optimal anesthesia is reached with a mixture of isoflurane and nitric oxide, however breathing needs to be monitored and eventually lowered to around 1-1.5%.

**Note 4.** If loose connective tissue is left in place there is a risk of the cyanoacrylate helmet to de-attach and fall ones the miniscope is connected, as it is heavy (~3g) in comparison just the baseplate, and thereby you might lose the whole preparation if this step is not done with great care.

**Note 5.** Bone residues are white in color and easy to spot at the edge of the drilled hole with the brain tissue as background color. Moreover, during the drilling of the cranial hole there might be small blood coagulates. These blood clots are usually naturally removed within 2-3 weeks, and should not be a major concern.

**Note 6.** We have found that leaving the lens to fall down due to gravity when the pump is turned off is not providing enough stability for applying the cyanoacrylic glue. Therefore, we have modified the protocol by Resendez *et al.,* by using the pipette tip to support the lens[21]. In detail, we lower the edge of the tip onto the lens, taking great care not to press down the lens (by visual inspection using the stereoscope), but holding it in place. In our experience, when we have not done this the imaging has been poorer, as the lens might not be exactly in the correct position/depth. Thereby you need to adjust the working distance more and might end up with just a faint blurry image.

**Note 7.** It is important to NOT completely cover the screws, as they will also anchor the miniscope baseplate later (section 3.7). Also, in our experience some cyanoacrylate glue is better suited for this procedure than others, for example Optibond FL 2, have much better adherence to the cranial surface than cheaper auto polymerizing acrylics.

**Note 8.** For prism lenses, the imaging is done lateral to the prism tip, so make sure that the coordinates are adjusted accordingly. Also, that the lens is inserted with the longer edge of the triangular prism tip facing the area to be imaged see Fig. 2.

**Note 9**. We have modified this step compared to other protocols as our prism lens is separate from the baseplate[25]. We simply use a standard 200µl pipette tip that we have cut to have a tip opening that can secure the 1 mm diameter prism lens. Now there is also the option of using 3D printed baseplates that are adapted for securing the 1mm prism lens[50].

**Note 10.** We administer anti-inflammatory drugs subcutaneously before the surgery and 5 days after, and this will also help to keep lenses clear after surgery[50].

**Note 11**. If during the implantation of the lens there is no bleeding or if it is contained, there is the possibility to check that indeed the lens has been positioned in a place where cells express genetically encoded fluorescent calcium indicators. After creating a fixed helmet on the bone, using the stereotaxic, position the miniscope on the top of the lens (in the case of prism lenses, use a relay lens attached to the miniscope (section 3.8). Check the image using acquisition software. To see cells or a fluorescent blur is a positive sign.

**Note 12.** We use a 3D printed miniscope holder[52].

**Note 13.** In the initial step it is good to apply a very small drop of glue on the side of the GRIN-lens as the glue will have to be broken to remove the miniscope later from the relay lens system.

**Note 14.** This section was developed by us, where we glue the two lenses together (with an air gap). Other protocols use the GRIN-lens pre attached to the minicope[21,25,31]. We have found that gluing the GRIN relay lens and the prism lens together provides better stability and makes focus adjustments easier as both lenses are pre-fixed to each other. This also makes the gluing of the baseplate easier, as any small touching by the baseplate will not de-align the lenses (as they are already glued together on the side).

**Note 15.** Some examples of software compatible with miniscope v3 are the MiniscopeControl[32,53], and Miniscope-DAQ-QT-software[33] for v4.

**Note 16.** To avoid missed frames due to computer limitations in processing and in memory, it is advisable to close all other running programs during video capture.

**Note 17.** Minian package from Denise Cai's lab: Minian was written in Python language and is suggested to use the Jupyter Notebook as the working environment[39]. This package has been tested using OSX, Linux, and Windows operating systems, and can be installed in systems with low computing resources. Calcium activity can be extracted using constrained non-negative matrix factorization[30] implemented in the Minian package from Denise Cai's lab[54]. MiniAn is an analysis pipeline and visualization tool inspired by both CaImAn and MIN1PIPE package specifically for Miniscope data. Minian has been validated to extract calcium events across different brain regions and cell types and includes resources for cross cell registration in different sessions. In practice, Minian provides an open-source calcium imaging analysis pipeline with interactive visualizations to explore parameters and validate results[39].

**Note 18.** PIMPN - Python Integrated Miniscope Pipeline Notebook: This package was developed to analyze calcium images and behavior videos directly in the cloud, through the Dropbox API for transferring miniscope data and the results[47]. This tool was written in Python (run in a jupyter Notebook environment), provides free access of powerful parallel computing resources provided by the use of the google colab, and is a very interesting start point for extracting calcium dynamics from miniscope raw data, as there is no need for installing programs, no need for download data on your local computer, is fully automatic and unsupervised, is fast and can be conveniently accessed by an internet web page and used by any computer or tablets. This tool uses CaImAn and NormCorre for motion correction as well as Constrained Non-Negative Matrix Factorization (CNMF/CNMFE) for source extraction and can also correlate the data with a trained deep learning model. At the present time this package has no official support, is compatible with previous versions of CaImAn, but it is still fully functional giving results as calcium dynamics and deconvoluted traces that can be saved in a matlab file and also hdf5 files. An updated version of this package that uses Google Drive instead of Dropbox API can be obtained upon request to the corresponding author.

**Note 19.** CellReg - Cell Registration package: This package was written in Matlab language, and using a probabilistic approach allows one to automatically track the same cells in multiple sessions and estimates the confidence of the record for each registered cell[48]. Calcium imaging data tested with this method produced estimated error rates <5%, and the registration is scalable for many sessions[36].

**Figure legends**

**FIGURE 1. Images illustrating lens implantation into the hippocampus. A)** Photo showing how two suture threads can stretch the skin apart to give better access to perform fast and clean surgery. Note that the treads are held by the stereotaxic device. **B)** Placement of anchor micro screws around the area of interest. **C)** Photo of the micro screws being screwed manually into the drilled holes in the bone without going through the bone. **D)** Photo of the cranial window position (arrow) relative to anchor screws. **E)** A bent syringe tip as a tool to remove the bone cap. **F)** The aspiration syringe and tubing. **G)** The adapted plastic tip with diameter to hold the GRIN lens with negative pressure from the vacuum pump (or manually inserted into the tip if the diameter is optimal for the lens. **H)** The GRIN lens being placed into the cranial window using the stereotaxic device. **I)** Note the height of the anchor screws in relation to the lens. **J)** Two knots (anterior and posterior, arrows) to avoid any further stretching of skin and visualization of how clear the silicon cap color is (to not be confused with bare bone).

**FIGURE 2. Schematic representation of the GRIN relay lens and prism lens.** Note that the prism should be lowered to acquire the image of the region of interest in the center when being implanted (see figure 1). Also note the air space between the GRIN lens and the prism lens when attaching the base plate (see figure 2).

**FIGURE 3. Attachment of baseplate for miniscope. A)** For miniscope V3 the GRIN-relay lens is attached to the bottom of the miniscope with a tiny drop of glue (see arrow) when the miniscope stands upside down. **B)** Example of a simplified 3D printed miniscope holder (orange) attached to a rod in the stereotaxic frame. **C)** Image illustrating the air space between the implanted prism lens and the miniscope attached GRIN-relay lens (seen as blue from the laser light, arrow). **D)** Image illustrating cyanoacrylate glue being applied to the side of the prism lens and the relay lens to hold them together. **E)** Image showing the lifted miniscope and the GRIN-relay lens/prism probe lens held together with glue around, but NOT in between lenses. **F)** Lowering of the miniscope with the base plate attached (by magnets). **G)** The miniscope and base plate lowered onto the GRIN-lens to optimal imaging is achieved. **H)** Image illustrating the cyanoacrylate glue holding the baseplate. Notice that the micro screws have been covered to anchor the baseplate, but the edges of the base plate are half exposed for the connection of the miniscope on the day of behavioral experiments.

**FIGURE 4. Representative images of a mouse wearing a miniscope device.**
Wild-type adult (4-5 month) mouse fitted with the UCLA miniscope V3.

**FIGURE 5. CalmAn extracted cells and analysis using PIMPN.**
**A)** Example image from a raw data video after down sampling. **B)** Associated shifts for motion correction as x-axis shifts (blue) and y-axis shifts (orange). **C)** The motion corrected template image. **D)** Correlation projection image (477 neurons from the mouse motor cortex). Value above ~0.4 indicates neurons. **E)** Spatial footprints. **F)** Time course of calcium activity traces for the first 50 neurons. **G)** Spike events and magnitude of spiking activity derived from one cell[55] in (F).

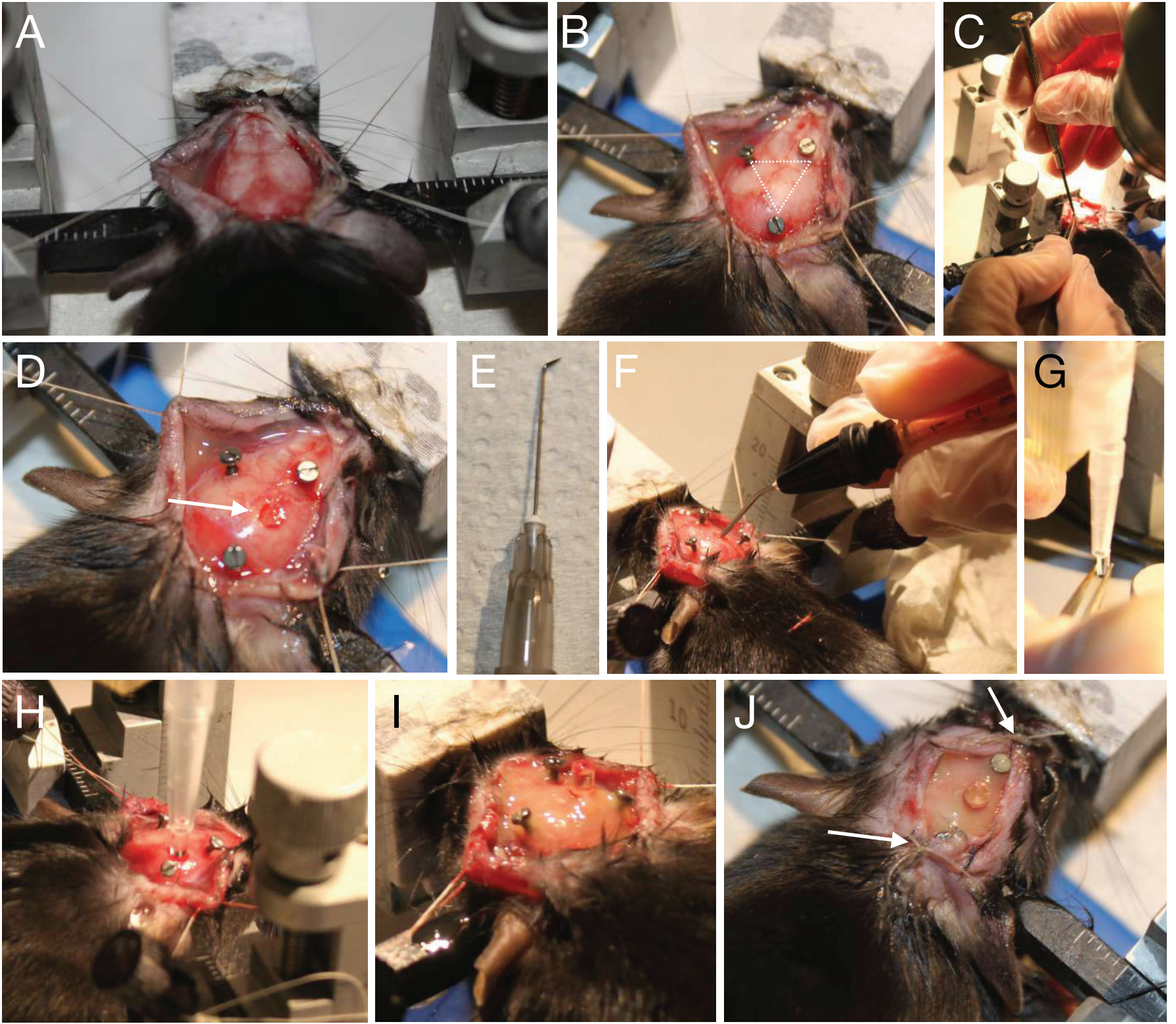

Figure 1

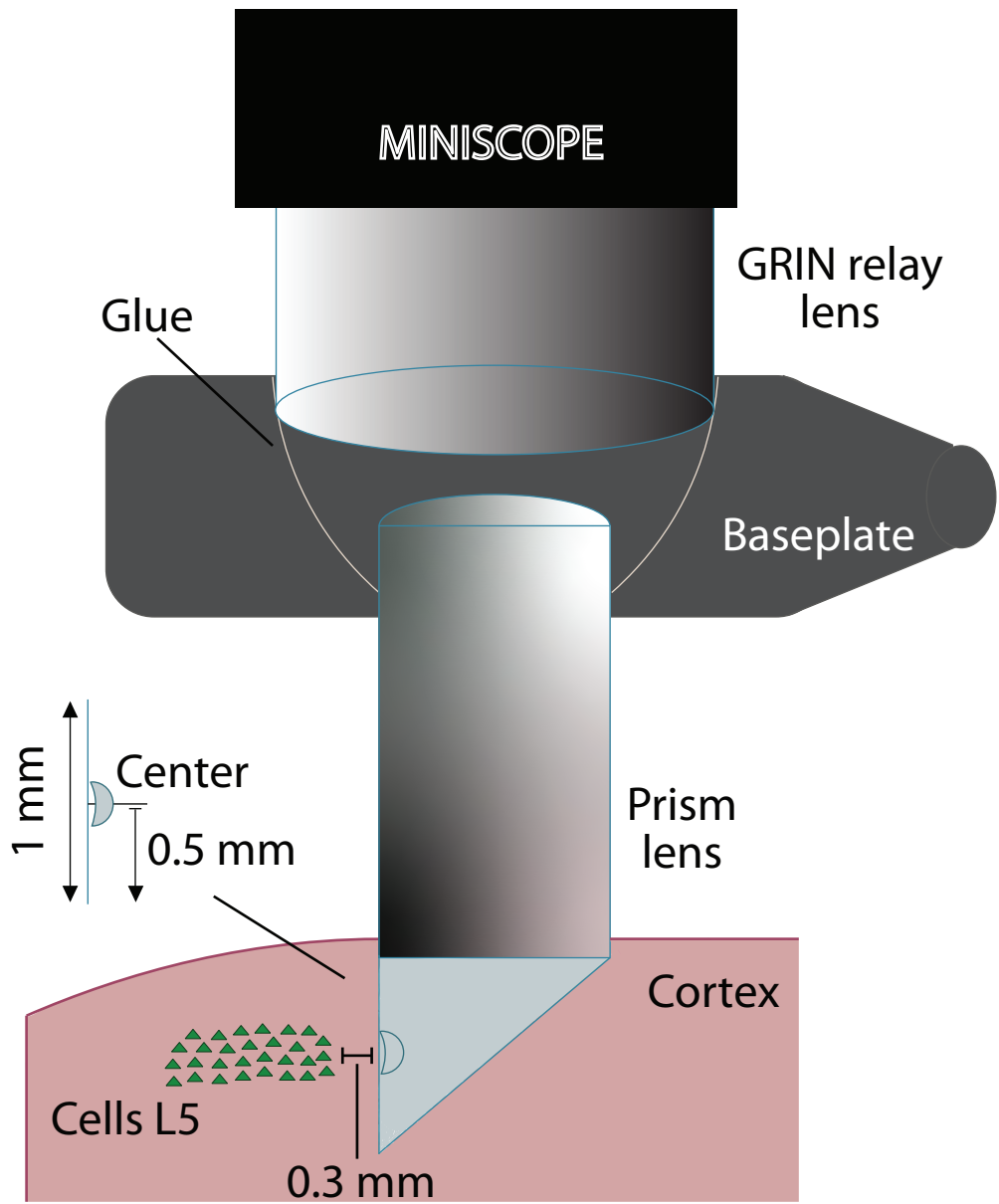

Figure 2

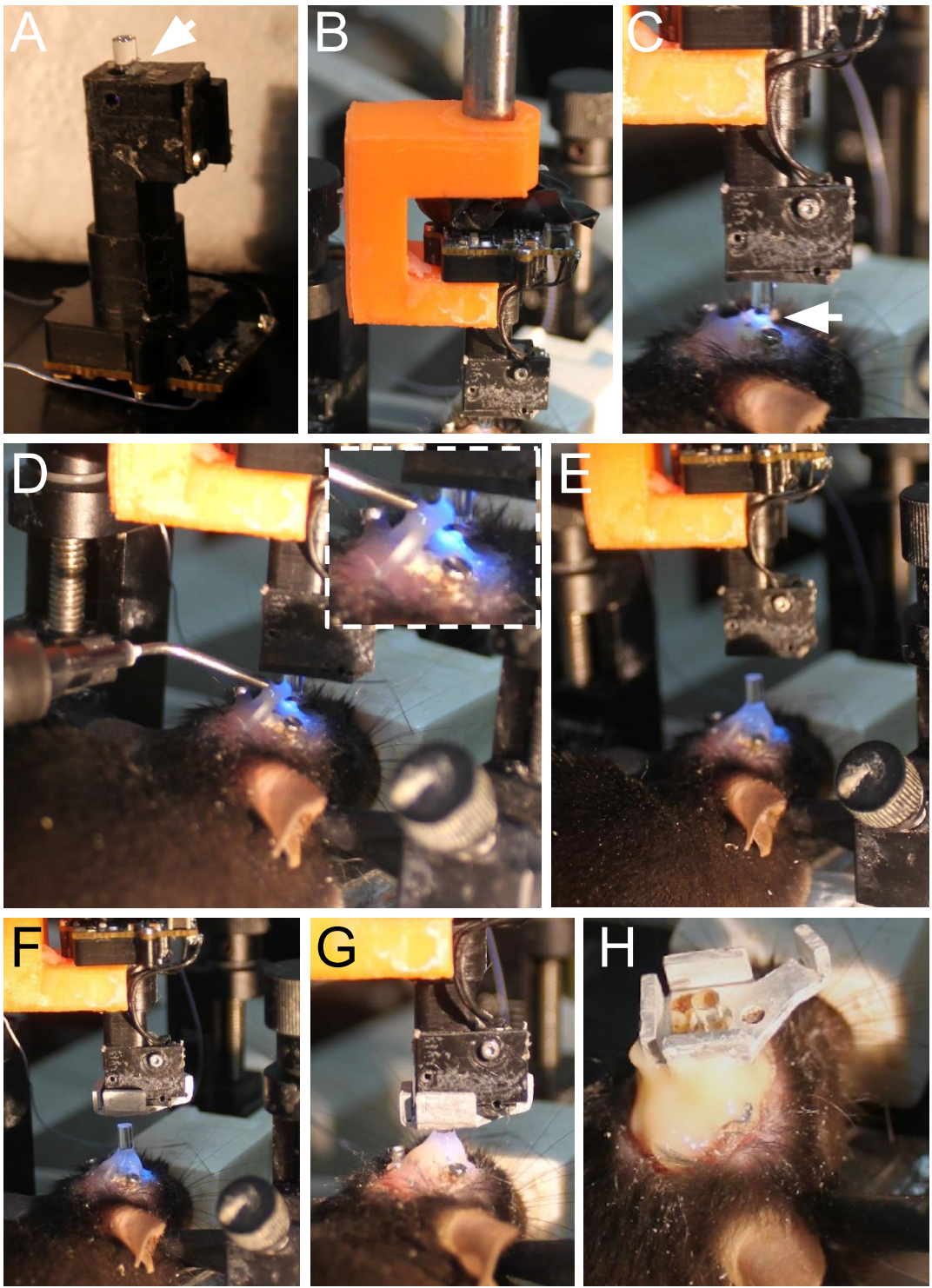

Figure 3

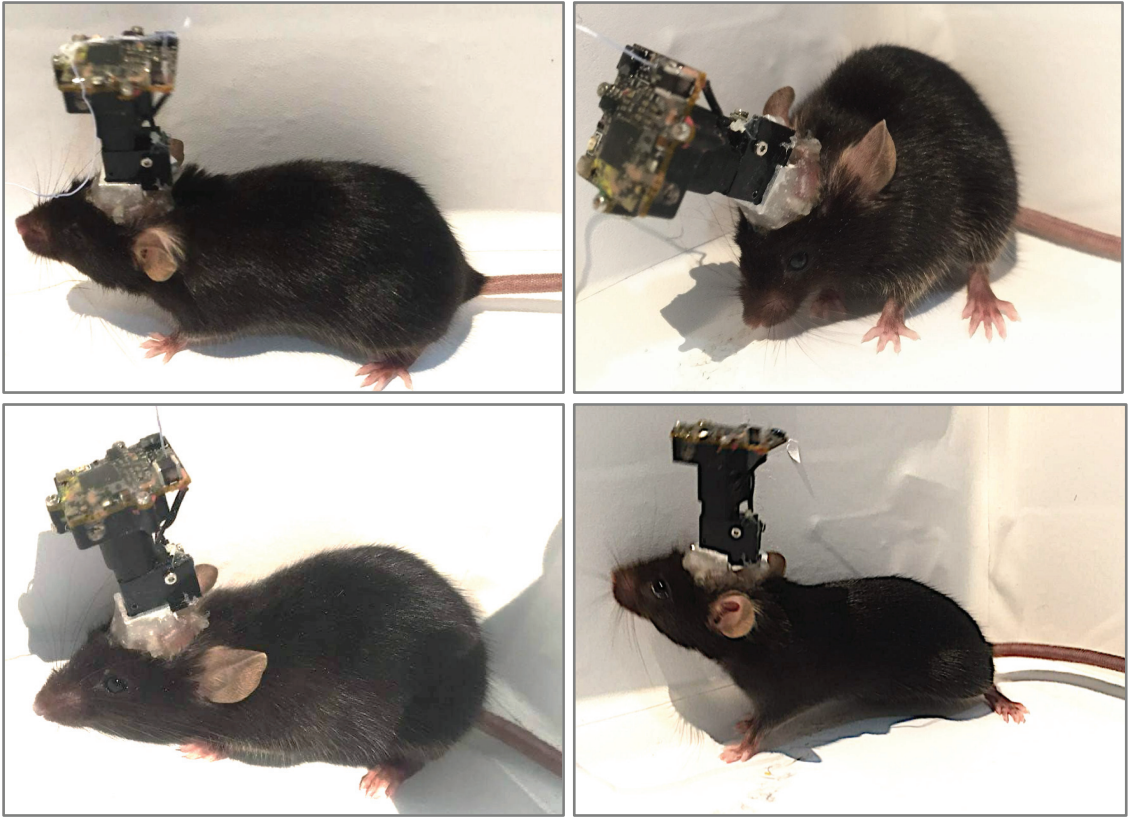

Figure 4

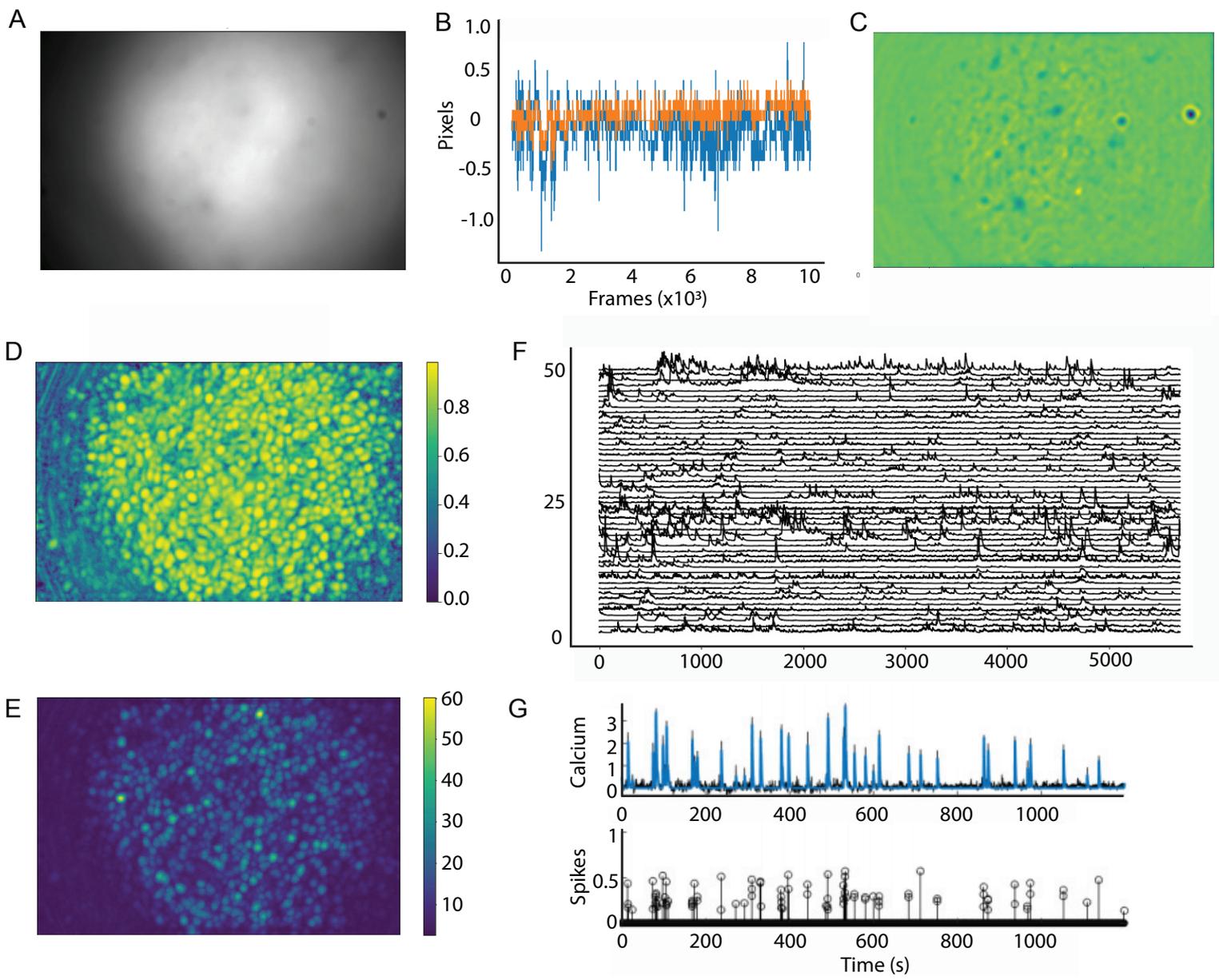

Figure 5